\documentclass[12pt]{article}
\usepackage{amsmath,amssymb,amsfonts,color,graphicx,cite,color}
\usepackage{latexsym}
\usepackage[normalem]{ulem}
\usepackage{epsfig,psfrag,rotating,soul}
\usepackage{rotfloat}
\usepackage{setspace}
\usepackage{orcidlink}
\input paperdef

\evensidemargin \oddsidemargin
\allowdisplaybreaks

\begin{document}
\thispagestyle{empty}

\def\thefootnote{\fnsymbol{footnote}}


\vspace{0.5cm}

\begin{center}

\begin{large}
\textbf{Interplay between $M_W$, $\Omega_{\rm CDM} h^2$, and $(g-2)_{\mu}$ in Flavor }
\\[2ex]
\textbf{Symmetry-Based Supersymmetric Models}
\end{large}

\vspace{1cm}

{\sc
S.~Israr$^{1}$%
\footnote{email: sajidisrar7@gmail.com},%
~M.~E.~G{\'o}mez\orcidlink{0000-0002-0137-0295},$^{2}$%
\footnote{email: mario.gomez@dfa.uhu.es},%
~M.~Rehman\orcidlink{0000-0002-1069-0637},$^{1}$%
\footnote{email: m.rehman@comsats.edu.pk}%
~and Y.~Arafat\orcidlink{0009-0000-6085-3740}$^{3}$%
\footnote{email: arafatfani439@gmail.com}%
}

\vspace*{.7cm}
{\sl
${}^1$Department of Physics, Comsats University Islamabad, 44000
  Islamabad, Pakistan \\[.1em] 
$^2$ Department of Applied Physics, University of Huelva, 21071 Huelva, Spain\\[.1em]
$^3$ Center For High Energy Physics, University of the Punjab, Quaid-e-Azam Campus, 54590 Lahore, Pakistan

}
\end{center}

\vspace*{0.1cm}

\begin{abstract}
\noindent
We study the phenomenological implications of the minimal supersymmetric standard model (MSSM) augmented by a non-abelian flavor symmetry labeled as sMSSM. Incorporating this flavor symmetry allows for a significant reduction in the original plethora of free parameters present in the MSSM, ultimately reducing them down to just seven in sMSSM. This reduction of free parameters is not achieved through ad hoc assumptions like in the constrained MSSM (CMSSM); rather, it is grounded in theoretical considerations. Our work focuses on exploring the interplay between the $W$ boson mass ($M_W$) predictions, the cold dark matter (CDM) relic abundance ($\Omega_{\rm CDM} h^2$), and the $(g-2)_{\mu}$ anomaly. We identified correlations among the theoretical parameters arising from this interplay, which can be complemented by experimental constraints such as the Higgs boson mass, B-physics observables, and charge and color breaking minima. Additionally, our investigations show that the $(g-2)_{\mu}$ discrepancy and the Planck bounds on $\Omega_{\rm CDM} h^2$ can be addressed within the sMSSM, but only in a very narrow region of the parameter space.
\end{abstract}

\def\thefootnote{\arabic{footnote}}
\setcounter{page}{0}
\setcounter{footnote}{0}

\newpage


\section{Introduction}
\label{sec:intro}

The Minimal Supersymmetric Standard Model (MSSM)~\cite{Fayet:1974pd,Fayet:1976et,Fayet:1977yc, Nilles:1983ge, Haber:1984rc,Barbieri:1987xf} is widely recognized as one of the most popular theories beyond the Standard Model (SM) \cite{Glashow:1961tr,Weinberg:1967tq, Salam:1968rm, Glashow:1970gm}. However, due to its extensive set of more than 105 free parameters, making meaningful phenomenological predictions becomes challenging. To address this issue, researchers have explored different approaches to reduce the number of free parameters. One such approach is the Constrained MSSM (CMSSM)\cite{Kane:1993td}, where specific assumptions are employed to simplify the model and bring down the number of free parameters to just five. It is important to note that these simplifications are not based on theoretical considerations but rather rely on ad hoc assumptions. 

While the CMSSM offers a highly idealized version, it is worth mentioning that the realistic supersymmetric (SUSY) models may encompass additional complexities and non-universalities in the soft SUSY-breaking (SSB) parameters. These intricacies can significantly impact collider searches, cosmology, and other experimental and theoretical investigations. Moreover, the experimental landscape does not currently favor the CMSSM, as recent findings from the CMS and ATLAS experiments have imposed rigorous limitations on the CMSSM parameter space\cite{Sekmen:2022vzu}. As a result, researchers are actively investigating alternative models, including the phenomenological MSSM and models based on symmetry principles. An example of an alternative model to the CMSSM is the Flavor Symmetry-Based MSSM (sMSSM)\cite{Babu:2014sga}, which incorporates a non-abelian flavor symmetry denoted as $H$. The number of free parameters in the sMSSM remains at a manageable level, with just two additional parameters compared to the CMSSM. 

Recently, the CDF collaboration has unveiled an intriguing indicator of potential new physics through its assessment of the mass of the $W$ boson\cite{CDF:2022hxs} displaying a significant deviation of approximately $7\sigma$ with the SM predictions~\cite{ParticleDataGroup:2022pth}. Apart from this discrepancy, there has also been considerable attention given to the long-standing anomaly in the $(g-2)_{\mu}$ value. The most recent assessment conducted by the Muon $g-2$ collaboration at Fermilab \cite{Muong-2:2023cdq,Muong-2:2021ojo}, when combined with the earlier findings from the Brookhaven E821 experiment \cite{Muong-2:2006rrc}, reveals a deviation of $5.1\sigma$ from the SM prediction\cite{Aoyama:2020ynm}.

Numerous studies in the existing literature have focused on exploring the phenomenological aspects of the sMSSM, including the $(g-2)_{\mu}$ anomaly and predictions for $\Omega_{\rm CDM} h^2$~\cite{Babu:2014lwa, Babu:2020ncc, Hussain:2017fbp}. In light of the recent CDF results and their implications for the $M_W$ world average, predictions for the $W$ boson mass have also become crucial. The interplay between $\Omega_{\rm CDM} h^2$, $(g-2)_{\mu}$, and $M_W$ can provide valuable insights for the MSSM~\cite{Bagnaschi:2022qhb}.

This work involves computing $\Omega_{\rm CDM} h^2$, $(g-2)_{\mu}$, and $M_W$ within the sMSSM frameworks. The main objective is to determine if the $(g-2)_{\mu}$ and $M_W$ anomalies, along with $\Omega_{\rm CDM} h^2$, can be explained within the sMSSM. Our analysis is conducted through the utilization of {\tt SARAH}~\cite{Staub:2009bi,Staub:2010jh,Staub:2012pb,Staub:2013tta,Staub:2015kfa}, {\tt SPheno}~\cite{Porod:2003um}, {\tt micrOMEGAs}\cite{Belanger:2006is,Belanger:2013oya,Barducci:2016pcb} and {\tt SARAH Scan and Plot} ({\tt SSP})~\cite{Staub:2011dp} setup. Initially, we employed {\tt SARAH} to generate the MSSM source code for {\tt SPheno}, facilitating subsequent tasks such as spectrum generation and the computation of low-energy observables like B-Physics observables, $\Delta \alpha_{\mu}$, and the mass of the $W$ boson using {\tt SPheno}. Moreover, the output from {\tt SPheno} was employed as input for {\tt micrOMEGAs} to evaluate the relic density of dark matter.

The paper is organized as follows: first, we present the main features of the sMSSM in \refse{sec:model_sMSSM}. Elaboration on the specifics of the calculations regarding low-energy observables like $M_W$ and new physics contributions to $(g-2)_{\mu}$ namely the $\Delta \alpha_{\mu}$ can be found in \refse{sec:CalcSetup}. The computational details and numerical results are presented in \refse{sec:NResults}. Our conclusions can be found in \refse{sec:conclusions}.  

\section{Flavor symmetry-based MSSM}
\label{sec:model_sMSSM}

The MSSM represents the most basic form of SUSY that can be constructed using the particle content of the SM. The overall configuration for the SSB parameters is given by~\cite{Fayet:1974pd,Fayet:1976et,Fayet:1977yc, Nilles:1983ge, Haber:1984rc,Barbieri:1987xf}
\begin{eqnarray}
\label{softbreaking}
-\cL_{\rm soft}&=&(m_{\tilde Q}^2)_i^j {\tilde q}_{L}^{\dagger i}
{\tilde q}_{Lj}
+(m_{\tilde u}^2)^i_j {\tilde u}_{Ri}^* {\tilde u}_{R}^j
+(m_{\tilde d}^2)^i_j {\tilde d}_{Ri}^* {\tilde d}_{R}^j
\nonumber \\
& &+(m_{\tilde L}^2)_i^j {\tilde l}_{L}^{\dagger i}{\tilde l}_{Lj}
+(m_{\tilde e}^2)^i_j {\tilde e}_{Ri}^* {\tilde e}_{R}^j
\nonumber \\
& &+{\tilde m}^2_{1}h_1^{\dagger} h_1
+{\tilde m}^2_{2}h_2^{\dagger} h_2
+(B_{\mu} h_1 h_2
+ {\rm h.c.})
\nonumber \\
& &+ ( A_d^{ij}h_1 {\tilde d}_{Ri}^*{\tilde q}_{Lj}
+A_u^{ij}h_2 {\tilde u}_{Ri}^*{\tilde q}_{Lj}
+A_l^{ij}h_1 {\tilde e}_{Ri}^*{\tilde l}_{Lj}
\nonumber \\
& & +\frac{1}{2}M_1 {\tilde B}_L {\tilde B}_L
+\frac{1}{2}M_2 {\tilde W}_L^a {\tilde W}_L^a
+\frac{1}{2}M_3 {\tilde G}^a {\tilde G}^a + {\rm h.c.}).
\end{eqnarray}
Here, the symbols $m_{\tilde Q}^2$ and $m_{\tilde L}^2$ represent $3 \times 3$ matrices in the family space, with indices $i$ and $j$ denoting generations, encapsulating the SSB masses for the left-handed squark doublets ${\tilde q}_{L}$ and slepton doublets ${\tilde l}_{L}$ associated with the $SU(2)$ symmetry. Similarly, the matrices $m_{\tilde u}^2$, $m_{\tilde d}^2$, and $m_{\tilde e}^2$ incorporate the soft masses for the right-handed up-type squarks ${\tilde u}_{R}$, down-type squarks ${\tilde d}_{R}$, and charged sleptons ${\tilde e}_{R}$, which are $SU(2)$ singlets. The matrices $A_u$, $A_d$, and $A_l$ are also $3 \times 3$ matrices, signifying the trilinear couplings concerning up-type squarks, down-type squarks, and charged sleptons, respectively. The parameter $\mu$ corresponds to the Higgs mixing, while ${\tilde m}_1$, ${\tilde m}_2$, and $B_{\mu}$ stand as the SSB factors are related to the Higgs sector. Within this context, $h_1$ and $h_2$ denote the two Higgs doublets. Lastly, the terms $M_1$, $M_2$, and $M_3$ define the mass parameters of the bino, wino, and gluino respectively. 


\refeq{softbreaking} encompasses over 105 free parameters, rendering the generation of viable phenomenological predictions within the context of the MSSM nearly impractical. However, this extensive list of parameters is significantly pruned down to a mere 5 within the CMSSM. This simplification is achieved by making the assumption that scalar masses are equivalent at the GUT scale, and likewise, the gaugino masses are equal as well. The foundation of the CMSSM finds its roots in minimal supergravity\cite{Chamseddine:1982jx,Barbieri:1982eh,Hall:1983iz}. However, this structure doesn't stem from theoretical considerations; instead, it relies on some ad hoc assumptions. Additionally, results from experiments at the LHC have constrained the parameter range of the CMSSM to a minimum extent. 

A category of supersymmetric models where the form of the SSB Lagrangian is determined solely by considerations of symmetry have been developed to overcome the limitations posed by the CMSSM. This framework is known as the flavor symmetry-based MSSM (sMSSM) which is defined with a SSB Lagrangian that satisfies two essential symmetry conditions. Firstly, the parameters are consistent with a grand unified symmetry such as  $\mathtt{SO}\left(10\right)$. Secondly, a non-abelian flavor symmetry $H$ operates on the three generations of particles, effectively suppressing excessive Flavor Changing Neutral Currents (FCNC) mediated by the SUSY particles.

Explicit examples of these models are constructed using flavor symmetries such as gauged $\mathtt{SU}\left(2\right)_{H}$ and $\mathtt{SO}\left(3\right)_{H},$ where the three generations of particles transform as \textbf{$\boldsymbol{2}+\boldsymbol{1}$} and \textbf{$\boldsymbol{3}$} representations respectively. In the framework of $\mathcal{\mathtt{SU}}\left(2\right)_{H},$ there is a straightforward solution for suppressing flavor-violating D-terms based on an interchange symmetry and the $\mathcal{\mathtt{SO}}\left(3\right)_{H}$ symmetry framework naturally avoid the D-term issues. Moreover, within the framework of the $H$, it is sufficient for only the first two generations to constitute a shared multiplet to address the SUSY flavor problem. Meanwhile, the third family can be treated as a singlet. Therefore, both \textbf{$\boldsymbol{2}+\boldsymbol{1}$} assignment of fermion fields under $H$ as well as \textbf{$\boldsymbol{3}$} assignment are equally valid.

Compatibility with $\mathtt{SO}\left(10\right)$ unified symmetry has major advantages from the standpoint of SSB. By
dropping the number of soft masses for sfermions from fifteen associated with the SM gauge symmetry (matching to the fifteen chiral multiplets of the SM) to just three, it significantly lowers the number of free parameters for these particles. In addition, it offers a symmetry-based justification for the gaugino masses unification, decreasing the free parameters from three in the SM to one. There are  two more free parameters in sMSSM compared to CMSSM making the total up to 7. These are  
\[
\{m_{0_{1,2}},m_{0_{3}}, M_{1/2},A_{0},\tb, \mu, m_{A} \}.
\]
where $m_{0_{1,2}}$ representing the SSB mass parameter for the first two families of sfermions, $m_{0_{3}}$ for the third generation, $M_{1/2}$ for the SSB gaugino mass, $A_{0}$ for the SSB trilinear coupling. The parameters $\tb$, $\mu$ and $M_{A}$ represents the ratio of the vacuum expectation values of the two Higgs doublets, bilinear Higgs mixing term and the mass of the $\cp$-odd Higgs boson, respectively.

\section{Calculation of low-energy observables}
\label{sec:CalcSetup}

\subsection{SUSY contributions to the mass of the $W$ boson}
The CDF collaboration has disclosed a fascinating signal hinting at the possibility of new physics beyond the SM as they examined the mass of the $W$ boson. Their measurement yields a value\cite{CDF:2022hxs} 
\[
M_{W}^{\rm CDF}=80.4335\pm 0.0094 \gev,
\] 
displaying a significant deviation of approximately $7\sigma$ from the prediction within the SM\cite{ParticleDataGroup:2022pth}, where 
\[
M_{W}^{\rm SM}=80.357\pm 0.006 \gev.
\]
Upon combining the previous results of the experiments like ATLAS and LHCb for $M_{W}$, the world average is determined as\cite{deBlas:2022hdk} 
\[
M_{W}^{\rm avg}=80.4133\pm0.0080 \gev,
\] 
thereby resulting in a deviation of about $6.5\sigma$. Despite thorough considerations of quantum corrections to the mass of the $W$ boson (as discussed in reference\cite{Lu:2022bgw} and references therein), there still remains an outstanding discrepancy of approximately $5.1\sigma$ when compared to the SM prediction. In light of these findings, it becomes plausible to attribute this deviation from the SM to the potential influence of new physics beyond the SM.

The $W$ boson mass is modified by the shifts in the electroweak parameter $\De\rho$ according to
\begin{equation}
\label{eq:precobs}
\De\MW \approx \frac{\MW}{2}\frac{\cw^2}{\cw^2 - \sw^2} \De\rho.\
\end{equation}
where $\cw$ ($\sw$) are the $\cos$ ($\sin$) of the Weinberg angle $\theta_W$. The parameter $\De\rho$ can be calculated by the relation
\begin{equation}
\De\rho = \frac{\Si_Z^{\text{T}}(0)}{\MZ^2} -
          \frac{\Si_W^{\text{T}}(0)}{\MW^2}
\label{eq:drho}
\end{equation} 
where $\Si_{Z,W}^{\text{T}}(0)$ denote the unrenormalized transverse parts of the $Z$- and $W$-boson 
self-energies at zero momentum. The top and bottom quarks give the dominant contributions to the $\De\rho$ in the SM which is given by 
\begin{equation}
\De\rho = \frac{3 G_\mu}{8 \sqrt{2} \pi^2} F_0(m_{b}^2,m_{t}^2)
\label{eq:drho-SM}
\end{equation} 
where $G_\mu$ represents the usual Fermi constant while $m_{b}$ and $m_{t}$ represent the masses of bottom and top quarks respectively. The $F_0(m_{1}^2,m_{2}^2)$ is defined as 
\begin{align}
F_{0}(m^2_1, m^2_2) &=
m^2_1+m^2_2-\frac{2m^2_1 m^2_2}{m^2_1-m^2_2} \ln\KL\frac{m^2_1}{m^2_2}\KR.
\label{Fun:f0}
\end{align}

\begin{figure*}[htb!]
\centering
\includegraphics[width=12cm,height=10cm,keepaspectratio]{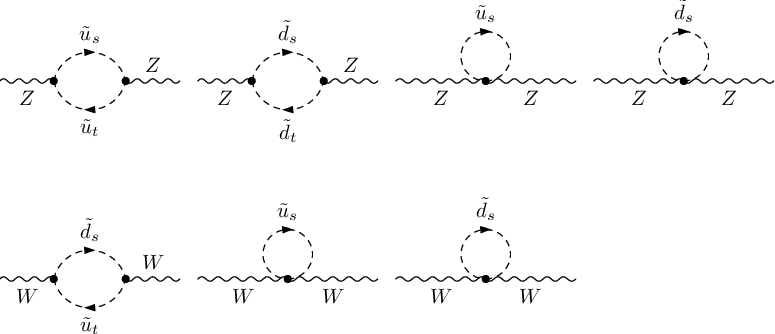}
\caption{One-loop self-energy Feynman diagrams for $Z$ and $W$~bosons. 
$\tilde u_{s,t}$ and $\tilde d_{s,t}$ represent the six mass eigenstates of up-type and down-type scalar quarks respectively.} 
\label{FeynDiagZWSelf}
\end{figure*}

In the SM, the contributions to $\De\rho$ originate from the
mass difference in one $SU(2)$ doublet. The picture is more intricate in the MSSM\cite{Heinemeyer:2004by,Heinemeyer:2004gx,Gomez:2015ila}. Representative Feynman diagrams for the scalar quarks contributions are shown in \reffi{FeynDiagZWSelf}. Here, only the  squarks contributions to $\De\rho$ are shown. However, $\De\rho$ also receive corrections from sleptons and other particles that exist within the MSSM framework. For our numerical evaluation, we have used {\tt SARAH} generated {\tt SPheno} code to calculate the $W$ boson mass which incorporate all these correction into account.


\subsection{SUSY contributions to $(g-2)_{\mu}$}
\label{sec:strategy0}

\begin{figure*}[htb!]
\centering
\includegraphics[width=12cm,height=10cm,keepaspectratio]{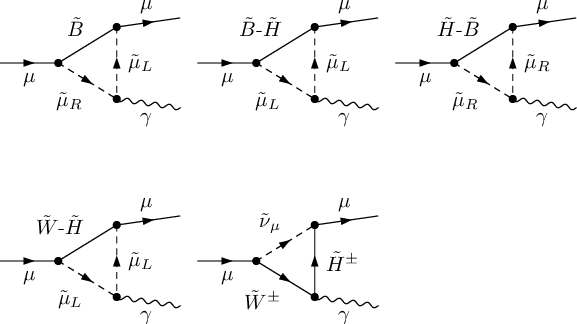}
\caption{Feynman diagrams for dominant MSSM contributions to $(g-2)_{\mu}$ originating from different neutralino and chargino species. The diagrams shown as ($\tilde B$-$\tilde H$), ($\tilde H$-$\tilde B$) and ($\tilde W$-$\tilde H$) represent mixing between neutralino species.}   
\label{diag(g-2)}
\end{figure*}

There has been considerable attention given to the long-standing anomaly in the $(g-2)_{\mu}$ value. The most recent assessment conducted by the Run II and Run III of Muon $g-2$ collaboration at Fermilab \cite{Muong-2:2023cdq}, when combined with the earlier findings from the same experiment\cite{Muong-2:2021ojo} and Brookhaven E821 experiment \cite{Muong-2:2006rrc}, reveals a deviation of $5.1\sigma$ \cite{Chakraborti:2023pis} from the Standard Model prediction\cite{Aoyama:2020ynm,Martin:2001st}
\begin{align}
\Delta \alpha_{\mu}=\alpha_{\mu}^{\rm exp}-\alpha_{\mu}^{\rm SM}= (24.9\pm 4.8) \times 10^{-10}.
\label{Eq:delAmu}
\end{align}
This discrepancy can be solved by considering the MSSM contributions to $(g-2)_{\mu}$. The dominant MSSM contributions originate from the sleptons, chargino and neutralino loops and are shown in \reffi{diag(g-2)}.

The corrections from the Feynman diagrams shown in \reffi{diag(g-2)} can be summarized as \cite{Moroi:1995yh, Stockinger:2006zn, Fargnoli:2013zia, Cho:2011rk,Gomez:2024dts} 

\begin{align}
\Delta \alpha_{\mu}^{\rm MSSM} &= \Delta \alpha_{\mu}^{{\tilde B}{\tilde \mu_L}{\tilde \mu_R}} 
+ \Delta \alpha_{\mu}^{({\tilde B}-{\tilde H}){\tilde \mu_L}} 
+ \Delta \alpha_{\mu}^{({\tilde H}-{\tilde B}){\tilde \mu_R}} \nonumber \\
&\quad+ \Delta \alpha_{\mu}^{({\tilde H}-{\tilde W}){\tilde \mu_L}} 
+ \Delta \alpha_{\mu}^{({\tilde W}-{\tilde H}){\tilde \nu_{\mu}}}
\label{eq:delAlphaMu}
\end{align}

where
\begin{align}
\label{delamu1}
\Delta \alpha_{\mu}^{{\tilde B}{\tilde \mu_L}{\tilde \mu_R}} & = \frac{g_Y^2}{8 \pi^2}\frac{m_{\mu}^2 \mu \tb}{M_1^3} F_b\KL\frac{m_{\tilde \mu_L}^2}{M^2_1},\frac{m_{\tilde \mu_R}^2}{M^2_1}\KR \\ 
\Delta \alpha_{\mu}^{({\tilde H}-{\tilde B}){\tilde \mu_R}} & = \frac{-g_Y^2}{8 \pi^2}\frac{m_{\mu}^2 M_1 \mu \tb}{m_{\tilde \mu_R}^4} F_b\KL\frac{M^2_1}{m_{\tilde \mu_R}^2},\frac{\mu^2}{m_{\tilde \mu_R}^2}\KR \\ 
\Delta \alpha_{\mu}^{({\tilde B}-{\tilde H}){\tilde \mu_L}} & = \frac{g_Y^2}{16 \pi^2}\frac{m_{\mu}^2 M_1 \mu \tb}{m_{\tilde \mu_L}^4} F_b\KL\frac{M^2_1}{m_{\tilde \mu_L}^2},\frac{\mu^2}{m_{\tilde \mu_L}^2}\KR \\ 
\Delta \alpha_{\mu}^{({\tilde H}-{\tilde W}){\tilde \mu_L}} & = \frac{-g^2}{16 \pi^2}\frac{m_{\mu}^2 M_2 \mu \tb}{m_{\tilde \mu_L}^4} F_b\KL\frac{M^2_2}{m_{\tilde \mu_L}^2},\frac{\mu^2}{m_{\tilde \mu_L}^2}\KR \\ 
\Delta \alpha_{\mu}^{({\tilde W}-{\tilde H}){\tilde \nu_{\mu}}} & = \frac{g^2}{8 \pi^2}\frac{m_{\mu}^2 M_2 \mu \tb}{m_{\tilde \nu_{\mu}}^4} F_a\KL\frac{M^2_2}{m_{\tilde \nu_{\mu}}^2},\frac{\mu^2}{m_{\tilde \nu_{\mu}}^2}\KR 
\label{delamu}
\end{align}

with $F_a\KL x,y \KR$ and $F_b\KL x,y \KR$ defined as

\begin{align*}
F_a\KL x,y \KR & = \frac{-1}{2 (x-y)}\KL \frac{(x-1)(x-3)+2 \ln x}{(x-1)^3}- \frac{(y-1)(y-3)+2 \ln y}{(y-1)^3}\KR \\ \nonumber
F_b\KL x,y \KR & = \frac{-1}{2 (x-y)}\KL \frac{(x-1)(x+1)-2x \ln x}{(x-1)^3} -\frac{(y-1)(y+1)-2 y \ln y}{(y-1)^3}\KR 
\end{align*} 

As can be seen from Eqs. (\ref{delamu1}-\ref{delamu}), all the contributions are proportional to $\mu \tb$ however the value of $\mu \tb$ is constrained by the vacuum stability bounds and electroweak symmetry breaking (EWSB).  

\section{Numerical Results}
\label{sec:NResults}

\subsection{Computational strategy}
\label{sec:strategy}

The following is a brief description of our computation workflow. With the help of Mathematica package {\tt SARAH}~\cite{Staub:2009bi,Staub:2010jh,Staub:2012pb,Staub:2013tta,Staub:2015kfa}, we first
created the MSSM source code for {\tt SPheno}~\cite{Porod:2003um}. {\tt SPheno} is a package that
numerically calculates SUSY mass spectra, decay rates of particles and various
low energy observables like $\De\rho$ and $\Delta \alpha_{\mu}$. Likewise, we generated the {\tt micrOMEGAs}\cite{Belanger:2006is,Belanger:2013oya,Barducci:2016pcb} source code using {\tt SARAH} to incorporate the constraints from the dark
matter observables into our analysis. These files are interfaced with {\tt SSP}~\cite{Staub:2011dp}, which is a Mathematica package that facilitates parameter scanning and plotting. For the sMSSM framework, the following parameter set was used for random scans:%
\[%
\begin{array}
[c]{ccc}%
 0 \leq& m_{0_{1,2}} & \leq5\ \tev\\
 0 \leq& m_{0_{3}} & \leq15\ \tev\\
 0 \leq& M_{1/2} & \leq2\ \tev\\
-3\leq&   A_{0}/m_{0_{3}} & \leq3\ \tev\\
0\leq & \mu & \leq 2 \tev \\ 
0\leq & M_{A} & \leq10\ \tev \\
1\leq & \tb & \leq 60.\ 
\end{array}
\]

These prescribed boundaries were applied at the GUT scale, with the exception of the parameters $\mu$ and $M_A$, which were instead defined at the SUSY scale. A private code written to calculate the charge and color breaking minima constraints (CCB) based on \citere{Beuria:2017gtf,Chattopadhyay:2019ycs} was also implemented in the {\tt SSP} package and data was generated in accordance with the following bounds on the low energy observables:%
\begin{equation}
\begin{aligned}
M_h & = 123-127~{\rm GeV},\\
m_{\tilde{g}} & \geq 2.1~{\rm TeV},\\
1.99\times 10^{-9} & \leq{\rm BR}(B_s \rightarrow \mu^+ \mu^-) \leq 3.43 \times10^{-9} \; (2\sigma), \\
3.02 \times 10^{-4} & \leq  {\rm BR}(B \rightarrow X_{s} \gamma)  \leq 3.62 \times 10^{-4} \; (2\sigma), \\
0.115 & \leq \Omega_{{\rm CDM}}h^{2} \leq 0.125 ~(5\sigma).
\end{aligned}
\label{eq:constraints}
\end{equation}

The current theoretical uncertainty in MSSM predictions for $\Mh$ stands at approximately the $2 \gev$ level ~\cite{Slavich:2020zjv}. Hence, we have opted for a range of $123 \gev \leq M_{h}  \leq 127\ \gev$. Regarding $M_W$, we have selected the $2\sigma$ range of $M_W^{\rm avg}$, while the values for ${\rm BR}(B\rightarrow X_{s}\gamma)$ and ${\rm BR} (B_{s}\rightarrow \mu^{+}\mu^{-})$ have been chosen at the $2\sigma$ level of their experimentally measured value\cite{ParticleDataGroup:2022pth,LHCb:2020zud}. The lower limit on $\Omega_{\rm CDM}$ \cite{Planck:2018vyg} may be disregarded as other particle species could also contribute to the DM relic abundance. Consequently, we exclusively consider points that align with the LSP neutralino relic density, ensuring it remains consistent with or lower than the Planck measurements. Nonetheless, we have examined the parameter space where both upper and lower limits could be satisfied.

\subsection{Constraints and Correlation between input parameters}
\label{sec:correlation}

In our effort to resolve the $W$ boson mass anomaly, the $(g-2)_{\mu}$ discrepancy, and the constraints imposed by $\Omega_{\rm CDM}h^{2}$ within the framework of the sMSSM, we meticulously examined the most appropriate parameter space. Our analysis revealed correlations between the input parameters and the various constraints on these parameters.

\begin{figure*}[ht!]
\begin{center}
\psfig{file=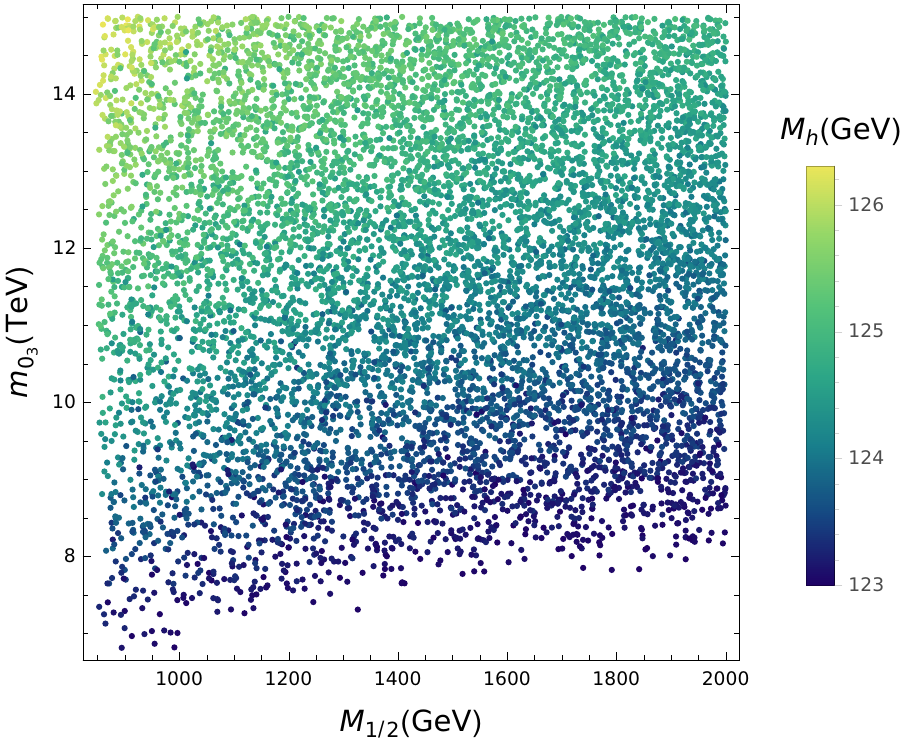  ,scale=0.50,angle=0,clip=}
\vspace{0.5cm}
\psfig{file=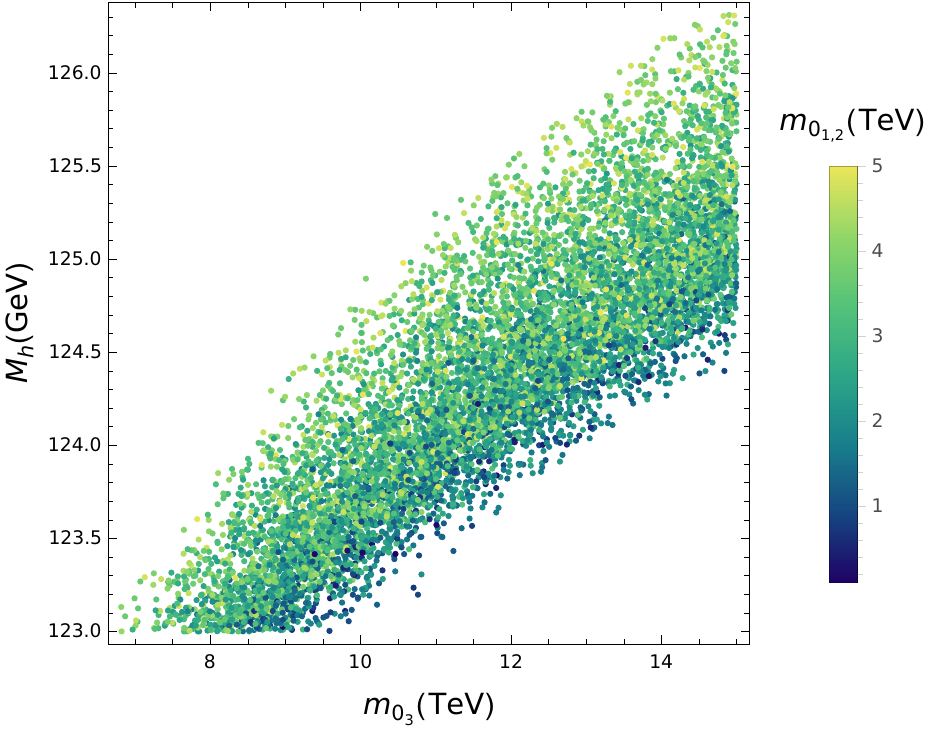,scale=0.50,angle=0,clip=} \\
\end{center}
\caption{The left plot displays the sMSSM's predictions for $M_h$ in the $M_{1/2}-m_{0_{3}}$ plane, with the color bar representing the value of $M_h$. The right plot illustrates $M_h$ versus $m_{0_{3}}$, with the color bar indicating the value of $m_{0_{1,2}}$.}
\label{fig:Mh0}
\end{figure*}

In the right plot of \reffi{fig:Mh0}, we present the sMSSM's predictions for $\Mh$ in the $M_{1/2}-m_{0_{3}}$ plane. The color bar indicates the value of $\Mh$. Experimental limits on the masses of squarks and gluinos constrain the $M_{1/2}$ parameter to be greater than $800 \gev$. It is also evident that $\Mh$ is highly sensitive to $m_{0_{3}}$ and increases with increasing $m_{0_{3}}$. However, no points are obtained for $m_{0_{3}} \lesssim 7 \tev$. This constraint primarily ensures that $M_{h}$ falls within the $123-127 \gev$ range. To clarify further, the left plot of \reffi{fig:Mh0} shows $\Mh$ versus $m_{0_{3}}$, with the color bar indicating the value of $m_{0_{1,2}}$.

\begin{figure*}[ht!]
\begin{center}
\psfig{file=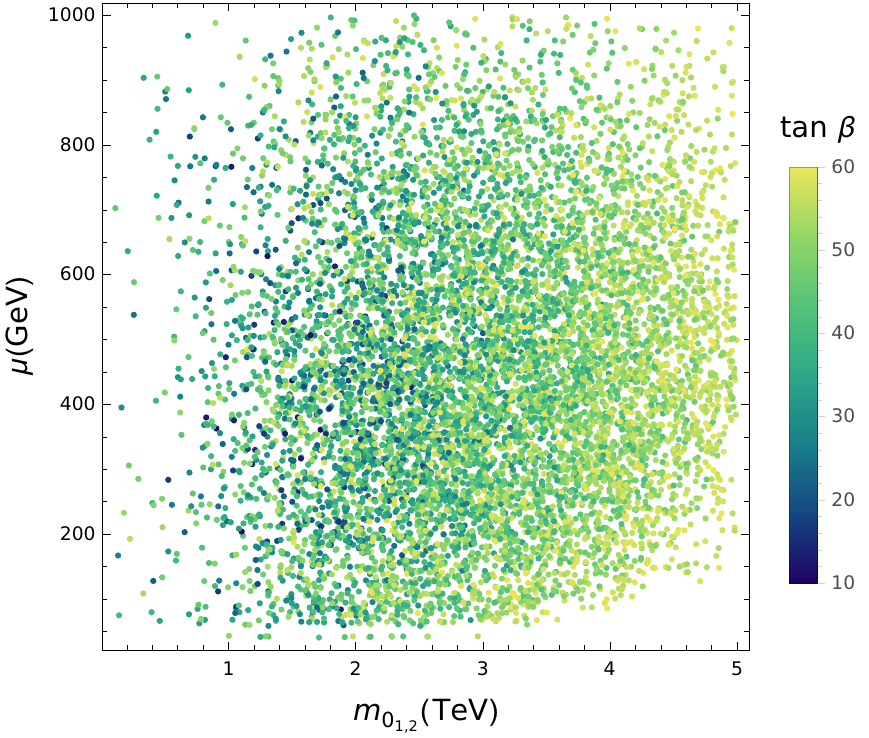  ,scale=0.50,angle=0,clip=}
\vspace{0.5cm}
\psfig{file=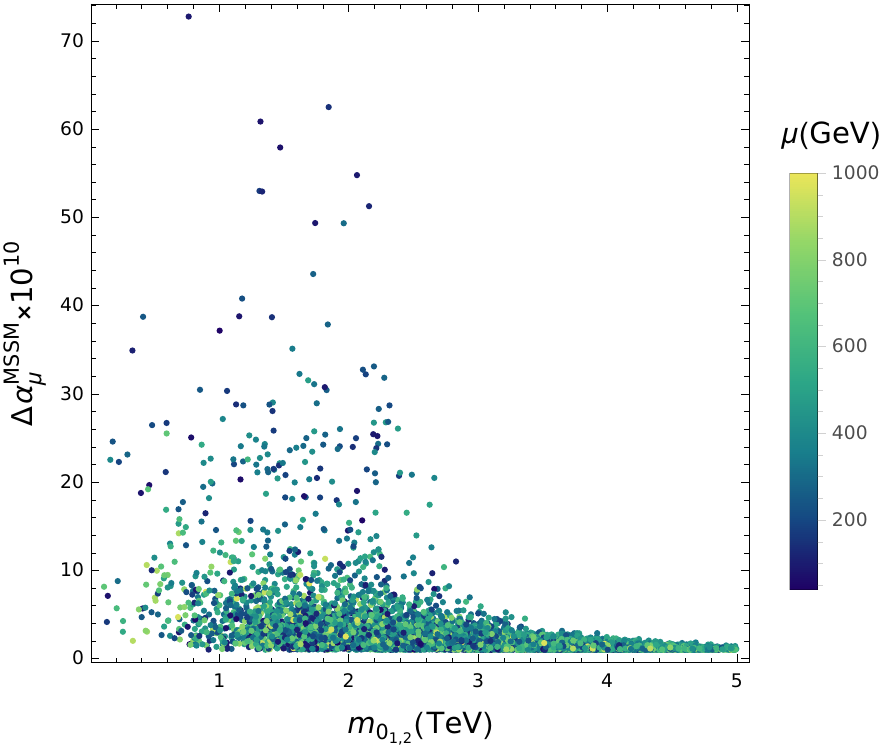,scale=0.50,angle=0,clip=} \\
\end{center}
\caption{The left plot displays the correlation between $\mu$, $m_{0_{1,2}}$, and $\tan\beta$, with the color bar indicating the value of $\tan\beta$. The right plot shows the correlation between $m_{0_{1,2}}$, $\Delta \alpha_{\mu}^{\rm MSSM}$, and $\mu$, with the color bar representing the value of $\mu$.}
\label{fig:m01mUEtanb}
\end{figure*}

In the left plot of \reffi{fig:m01mUEtanb}, we show the correlation between $\mu$, $m_{0_{1,2}}$ and $\tan\beta$. We observe that $m_{0_{1,2}}$ values increase with $\tan\beta$, with the highest $m_{0_{1,2}}$ values corresponding to extreme $\tan\beta$ values. Notably, there are no points for $\tan\beta < 10$, and $m_{0_{1,2}} < 3.5 \tev$  for $\tan\beta < 30$. Higher values of $m_{0_{1,2}}$ (above $3.5 \tev$) are only possible for $\tan\beta > 30$. This is likely related to electroweak symmetry breaking.
In the right plot of \reffi{fig:m01mUEtanb}, the correlation between $m_{0_{1,2}}$, $\Delta \alpha_{\mu}^{\rm MSSM}$ and $\mu$ is presented. As can be observed, the required values for $\Delta \alpha_{\mu}^{\rm MSSM}$ can only be obtained for $m_{0_{1,2}} < 3 \tev$.

\begin{figure*}[ht!]
\begin{center}
\psfig{file=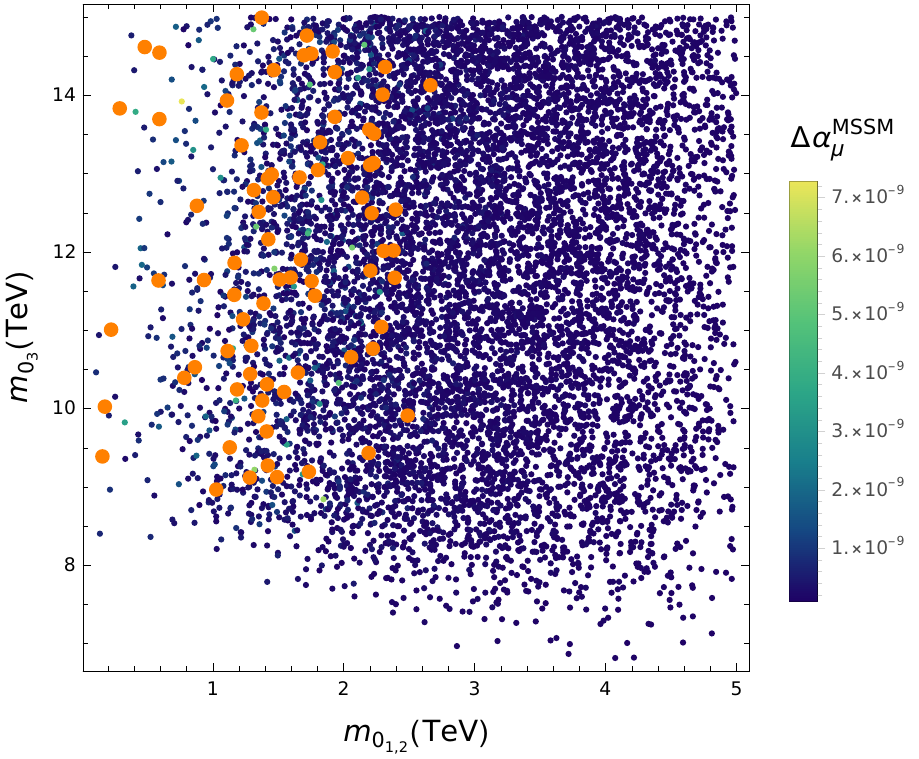  ,scale=0.50,angle=0,clip=}
\vspace{0.5cm}
\psfig{file=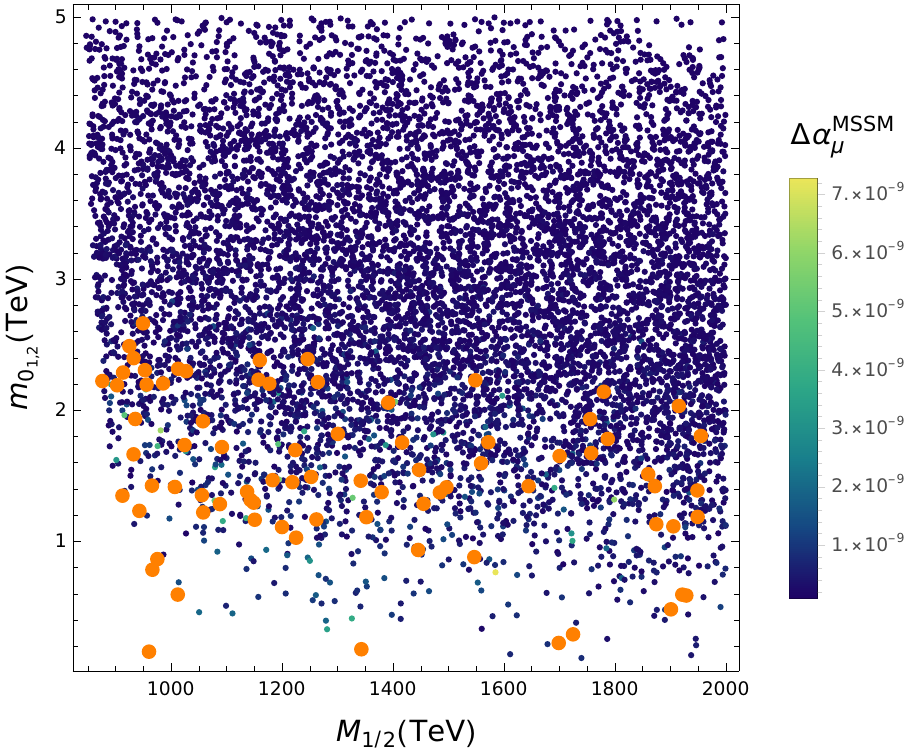,scale=0.50,angle=0,clip=} \\
\end{center}
\caption{The sMSSM's predictions for $\Delta \alpha_{\mu}^{\rm MSSM}$ in the $m_{0_{1,2}}-m_{0_{3}}$ (left plot) and $M_{1/2}-m_{0_{3}}$ (right plot) plane, with the color bar representing the value of $\Delta \alpha_{\mu}^{\rm MSSM}$. The orange points in the left plot represent locations where $\Delta \alpha_{\mu}^{\rm MSSM}$ falls within the $1 \sigma$ range required to address the $(g-2)_{\mu}$ discrepancy.}
\label{fig:GMm01m03}
\end{figure*} 

In summary, the experimental constraints on $\Mh$ and $m_{\tilde g}$, the upper limit on $\Omega_{\rm CDM}h^{2}$, and the $(g-2)_{\mu}$ anomaly further narrow down the parameter space of the sMSSM to: 
\[%
\begin{array}
[c]{ccc}%
 0 \leq& m_{0_{1,2}} & \leq 3\ \tev\\
 7 \leq& m_{0_{3}} & \leq15\ \tev\\
 0.8 \leq& M_{1/2} & \leq2\ \tev\\
0\leq & \mu & \leq 570 \gev \\ 
\end{array}
\]
The constraints on $m_{0_{1,2}}$ and $\mu$ originate from considerations of $(g-2)_{\mu}$, while the constraints on $m_{0_{3}}$ and $M_{1/2}$ are due to $M_h$ and $m_{\tilde g}$, respectively.  

\subsection{$\Delta \alpha_{\mu}$ and $\Omega_{\rm CDM}h^{2}$ in the sMSSM}

\begin{figure*}[ht!]
\begin{center}
\psfig{file=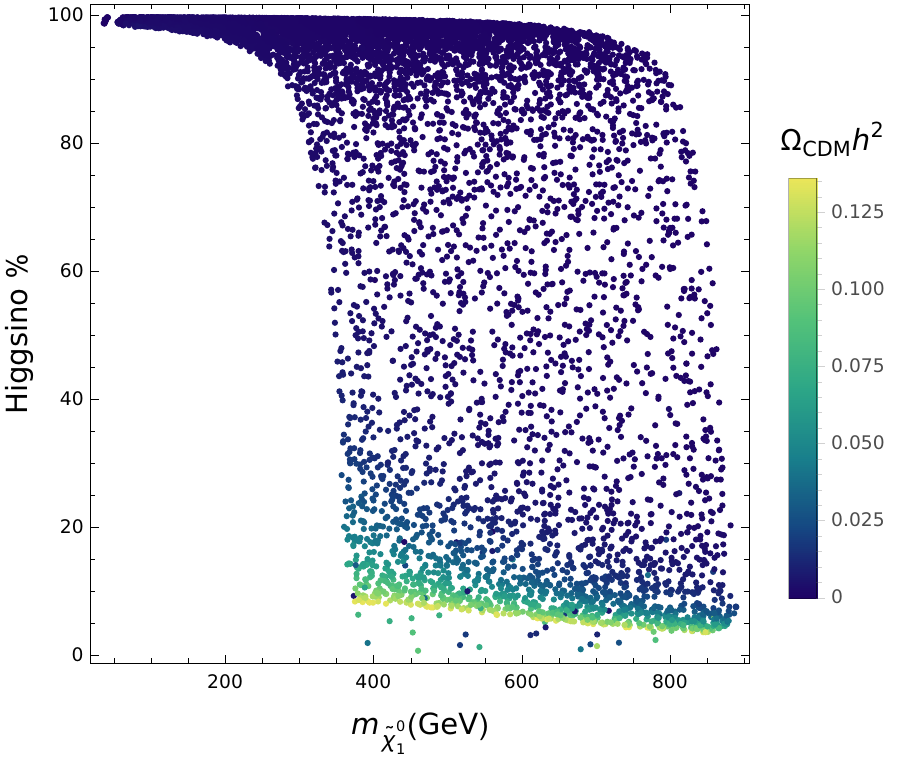  ,scale=0.50,angle=0,clip=}
\vspace{0.5cm}
\psfig{file=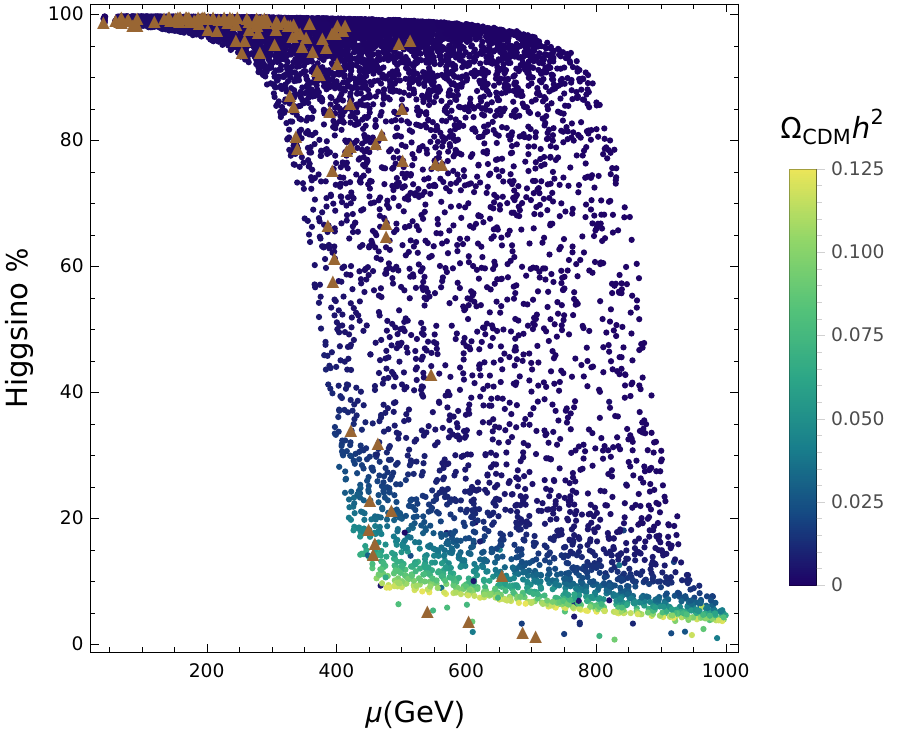,scale=0.50,angle=0,clip=} \\
\end{center}
\caption{Higgsino percentage versus $m_{\tilde{\chi}_1^0}$ (left plot) and $\mu$ (right plot), with the color bar indicating the value of $\Omega_{\rm CDM}h^{2}$.}
\label{fig:chi-higgsino-dm}
\end{figure*}

In this section, we discuss the predictions for $\Omega_{\rm CDM}h^{2}$ within the sMSSM and the resulting constraints imposed on the sMSSM parameter space by $\Omega_{\rm CDM}h^{2}$. Generally, we found that points contributing significantly to $M_W$ and satisfying the condition $\Omega_{\rm CDM} h^{2} \leq 0.125$ require a $\mu$ value below $1 \tev$. For these points, the lightest supersymmetric particle (LSP) is a neutralino, which is a mixture of higgsino and bino components. To characterize the higgsino component of the LSP, we consider the lightest neutralino as follows:
\begin{equation}
    \chi=N_{11} \tilde{B}+N_{12} \tilde{W}+ N_{13}\tilde{H}_1+N_{14}\tilde{H}_2.
\end{equation}
We characterize the higgsino composition as $\sqrt{N_{13}^2 + N_{14}^2} \times 100$. In the left (right) panel of \reffi{fig:chi-higgsino-dm}, we show the higgsino component of the LSP versus $m_{\tilde{\chi}_1^0}$ ($\mu$), with the color bar representing the value of $\Omega_{\rm CDM}h^{2}$. It is evident that the higgsino component is significant for most of the displayed points. Additionally, we observe that $\Omega_{\rm CDM}h^{2}$ increases as the higgsino component decreases. Throughout our entire parameter space, we find that the chargino is the Next to LSP (NLSP). 

\begin{figure*}[ht!]
\begin{center}
\psfig{file=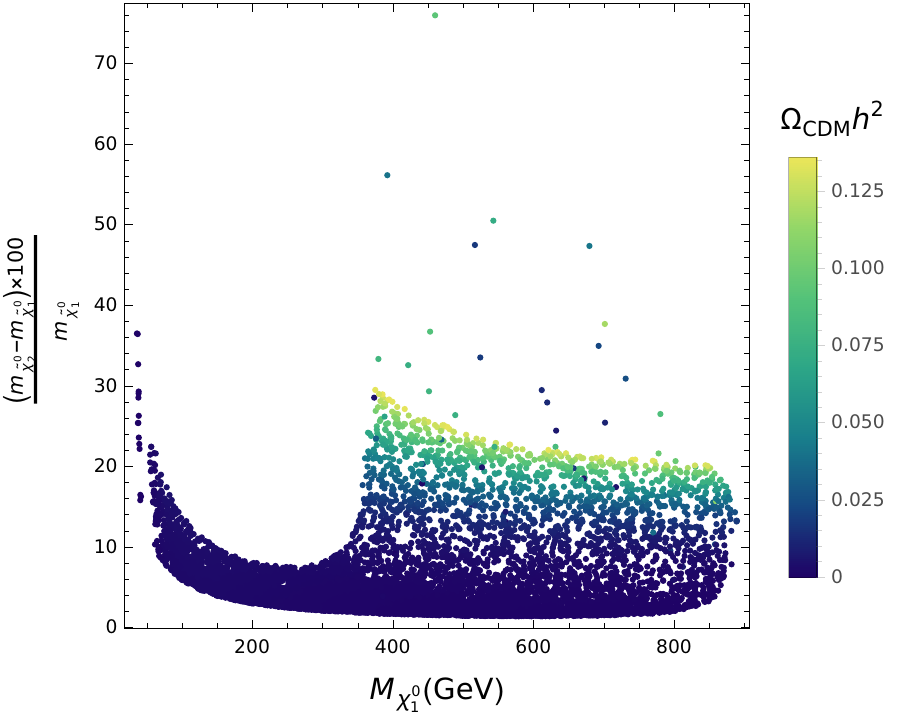  ,scale=0.50,angle=0,clip=}
\vspace{0.5cm}
\psfig{file=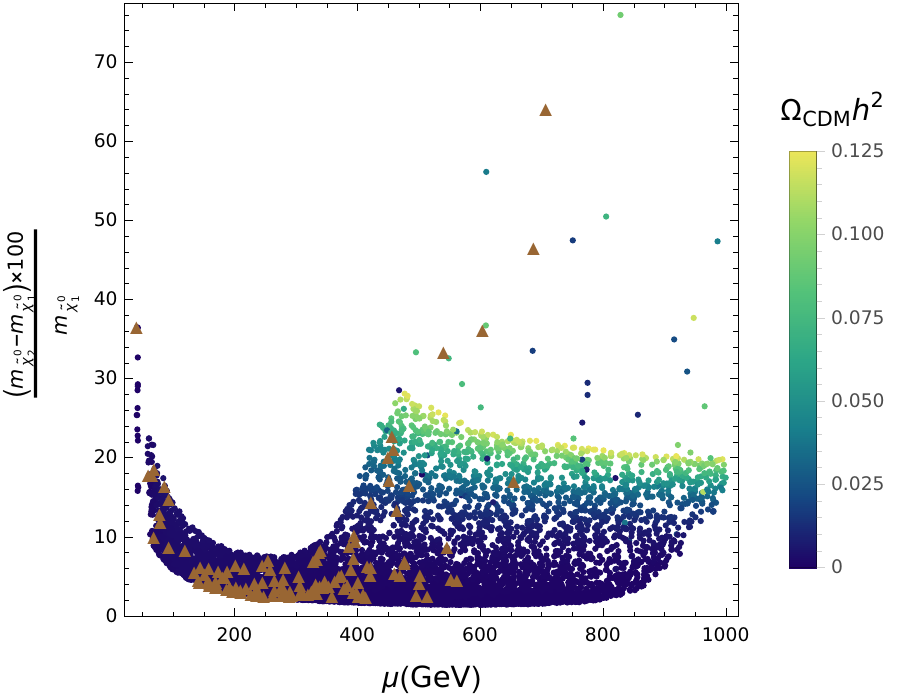,scale=0.50,angle=0,clip=} \\
\end{center}
\caption{Relative ratio $(m_{\tilde{\chi}_2^0}-m_{\tilde{\chi}_1^0}) \times 100/m_{\tilde{\chi}_1^0}$ verses $m_{\tilde{\chi}_1^0}$ (left plot) and $\mu$ (right plot) while the color bar depicts the value of $\Omega_{\rm CDM}h^{2}$.}
\label{fig:chi-relratio-dm}
\end{figure*}

In the left (right) panel of \reffi{fig:chi-relratio-dm}, we display the relative ratio  $(m_{\tilde{\chi}_2^0}-m_{\tilde{\chi}_1^0}) \times 100/m_{\tilde{\chi}_1^0}$ verses $m_{\tilde{\chi}_1^0}$  ($\mu$) while the color bar depicts the value of $\Omega_{\rm CDM}h^{2}$. It can be observed that models with neutralinos having less than 20\% Higgsino component can explain the dark matter relic density. In most of these models, the neutralinos can coannihilate with charginos, resulting in a relic density within the Planck bounds when the mass difference is between 20-30\%. We find only a few points with a Bino LSP that satisfy the Planck bounds, primarily due to resonances in the annihilation channels ($m_A \sim 2 \cdot m_\chi$). From \reffi{fig:chi-relratio-dm}, we can also see that achieving $\Omega_{\rm CDM} h^2$ in the range $0.115 < \Omega_{\rm CDM} h^2 < 0.125$ requires $m_{\tilde{\chi}_1^0}$ to be greater than $350 \gev$ and $\mu$ to be greater than $450 \gev$.

\begin{figure*}[ht!]
\begin{center}
\psfig{file=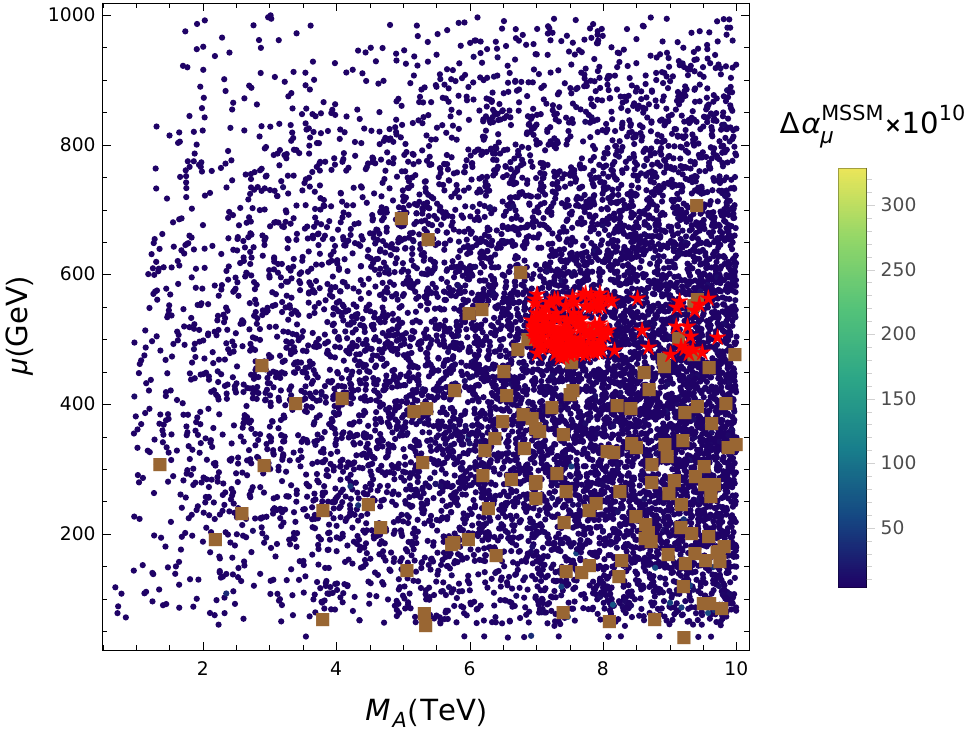  ,scale=0.50,angle=0,clip=}
\vspace{0.5cm}
\psfig{file=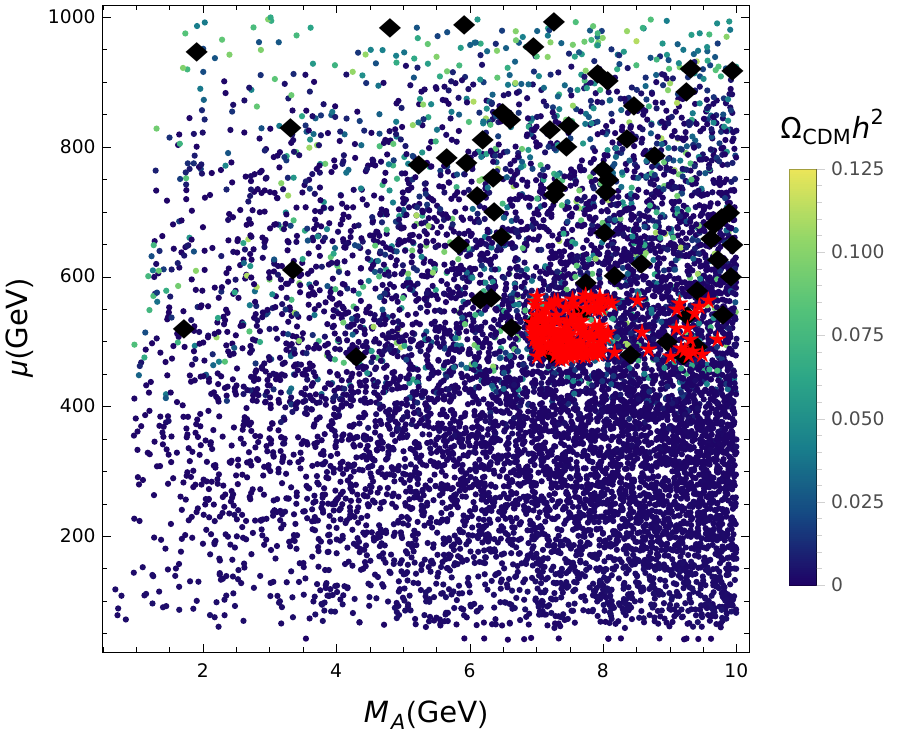,scale=0.50,angle=0,clip=} \\
\end{center}
\caption{The sMSSM predictions for $\Delta \alpha_{\mu}^{\rm MSSM}$ (shown in the left plot) and $\Omega_{\rm CDM}h^{2}$ (displayed in the right plot) shown in the $M_{A}-\mu$ plane. The color bar indicates the values of $\Delta \alpha_{\mu}^{\rm MSSM}$ (left plot) and $\Omega_{\rm CDM}h^{2}$ (right plot).}
\label{fig:MUEMA0-sMSSM}
\end{figure*}

In \reffi{fig:MUEMA0-sMSSM}, we present predictions for $\Delta \alpha_{\mu}^{\rm MSSM}$ (left plot) and $\Omega_{\rm CDM}h^{2}$ (right plot) in the $M_{A}-\mu$ plane. The color bar indicates the values of $\Delta \alpha_{\mu}^{\rm MSSM}$ (left plot) and $\Omega_{\rm CDM}h^{2}$ (right plot). The brown squares in the left plot represent locations where $\Delta \alpha_{\mu}^{\rm MSSM}$ falls within the $2 \sigma$ range required to address the $(g-2)_{\mu}$ discrepancy, while the black squares in the right plot correspond to $\Omega_{\rm CDM}h^{2}$ values within the $5 \sigma$ range of the Planck measurement. Notably, most of the brown points lie in the region where $\mu \leq 450 \gev$, whereas most of the black points are situated where $\mu > 450 \gev$. This discrepancy makes it challenging to simultaneously account for the $(g-2)_{\mu}$ discrepancy while adhering to the upper and lower limits of $\Omega_{\rm CDM}h^{2}$. The region where both $(g-2)_{\mu}$ and $\Omega_{\rm CDM}h^{2}$ can be satisfied is quite narrow, specifically $470 \gev \lesssim \mu \lesssim 570 \gev$ and $7 \tev \lesssim M_A $. In \refta{tab:Benchmark-Points-sMSSM}, we present three benchmark points that satisfy $\Mh$, $M_W$, $(g-2)_{\mu}$, and $\Omega_{\rm CDM}h^{2}$ constraints.

\begin{table}[ht]
\centering
\begin{tabular}{lccr}
\hline\hline
Parameter & $P_1$ & $P_2$ & $P_3$ \\ 
\hline \\ [-5pt]
$m_{0_{1,2}}$ & $1.774 \tev$ & $2.143 \tev$ & $1.835 \tev$ \\ [5pt]
$m_{0_{3}}$ & $14.844 \tev$ & $13.58 \tev$ & $10.317 \tev$ \\ [5pt]
$M_{1/2}$ & $900 \gev$ & $907 \gev$ & $1070 \gev$ \\ [5pt]
$\tan \beta $ & $53.012$ & $54.26$ & $56.14$ \\ [5pt]
$A_{0}$ & $-170.8 \gev$ & $-207 \gev$ & $-209 \gev$ \\ [5pt]
$M_{A}$ & $7.41 \tev$ & $9.286 \tev$ & $7.915 \tev$ \\ [5pt]
$\mu $ & $472.9 \gev$ & $477.6 \gev$ & $555 \gev$ \\ [5pt]
\hline \\ [-5pt]
$M_{h}$ & $125.24 \gev$ & $124.9 \gev$ & $123.63 \gev$ \\ [5pt]
$M_{W}$ & $80.397 \gev$ & $80.398 \gev$ & $80.398 \gev$ \\ [5pt]
$\Delta \alpha _{\mu }^{\text{MSSM}}$ & $20.45\times 10^{-10}$ & $23.08\times 10^{-10}$ & $20.67\times 10^{-10}$ \\ [5pt]
$\Omega _{\text{CDM}}h^{2}$ & $0.119$ & $0.121$ & $0.122$ \\  [5pt]
\hline\hline
\end{tabular}
\caption{Three benchmark points in the sMSSM that simultaneously satisfy the $\Mh$, $M_W$, $(g-2)_{\mu}$, and $\Omega_{\rm CDM}h^{2}$ constraints.}
\label{tab:Benchmark-Points-sMSSM}
\end{table}

\section{Conclusions}
\label{sec:conclusions}

 The flavor symmetry-based minimal supersymmetric standard model (sMSSM) is proposed as an alternative model to the constrained MSSM (CMSSM), which faces challenges in explaining experimental results from the LHC. This model utilizes non-abelian flavor symmetry at the GUT scale to constrain the number of free parameters, requiring only 7 in total. 

In this paper, we have investigated the predictions for the $W$ boson mass ($M_W$) and muon's anomalous magnetic moment $(g-2)_{\mu}$, as well as the dark matter relic abundance $\Omega_{\rm CDM}h^{2}$, within the sMSSM. In order to perform the calculations we generated the {\tt SPheno} and {\tt micrOMEGAs} source code for MSSM using Mathematica package {\tt SARAH}. The {\tt SPheno} and {\tt micrOMEGAs} codes were then used in the {\tt SSP} setup to generate numerical results for the MSSM particle spectra as well as the low energy observables like $M_W$, $\Delta \alpha_{\mu}$ and $\Omega_{\rm CDM} h^2$. For our numerical analysis, we randomly scanned the sMSSM free parameters while respecting the constraints from Higgs boson mass $M_h$, B-Physics observables (BPO), experimental limit on the gluino mass and DM relic density constraints.  

The $(g-2)_{\mu}$ anomaly indicates a preference for lower values of $m_{0_{1,2}}$, setting an upper limit of approximately $3 \tev$ on this parameter. To satisfy the experimental value of $M_h$, $m_{0_{3}}$ needs to exceed approximately $7 \tev$, ensuring $M_h$ falls within the range of $123-127 \gev$. The Higgs mixing parameter $\mu$ was varied from $0 \tev$ to $2 \tev$, but it is constrained by the upper limit on $\Omega_{\rm CDM}h^{2}$ to be below $1 \tev$, while $M_W^{\rm avg}$ also favors smaller values for $\mu$. The parameter $M_{1/2}$ was explored within the range of $0 \tev$ to $2 \tev$. Experimental constraints on the masses of gluino and scalar quarks require $M_{1/2}$ to exceed $800 \gev$, slightly higher than the previously reported lower limit of $700 \gev$ found in the literature. Additionally, for the sMSSM, achieving $\Omega_{\rm CDM} h^2$ in the range $0.115 < \Omega_{\rm CDM} h^2 < 0.125$ necessitates $m_{\tilde{\chi}_1^0}$ and $\mu$ to be greater than $350 \gev$ and $450 \gev$, respectively.  

In the sMSSM framework, addressing the $(g-2)_{\mu}$ anomaly prefers $\mu$ to be less than $450 \gev$, posing challenges in simultaneously satisfying Planck's measurement for $\Omega_{\rm CDM} h^2$. Our investigations indicate that the $(g-2)_{\mu}$ discrepancy and the Planck constraints on $\Omega_{\rm CDM} h^2$ can be met within the sMSSM, but though only within a very limited range of the parameter space, specifically $470 \gev \lesssim \mu \lesssim 570 \gev$ and $7 \tev \lesssim M_A $

\subsection*{Acknowledgments}

The research of M.~E.~G. is supported  by the Spanish Ministerio de Ciencia e Innovaci{\'o}n, under grant PID2022-140440NB-C22. The research of M.~R. is supported  by Higher Education Commission, Pakistan under NRPU grant 20-15867/NRPU/R\&D/HEC/2021.


\newpage
\pagebreak

\end{document}